\documentclass[seceq]{ptptex}
\usepackage{graphics}
\newcommand{\beq}{\begin{equation}}
\newcommand{\eeq}{\end{equation}}

\pubinfo{Vol. 127, No. 3, March 2012}
\title{Ghost Interference and Quantum Erasure}

\author{Pravabati \textsc{Chingangbam}$^1$ \footnote{E-mail: prava@iiap.res.in} and
Tabish \textsc{Qureshi}$^2$\footnote{E-mail: tabish@ctp-jamia.res.in}}

\inst{$^1$Indian Institute of Astrophysics, Koramangala, Bangalore-560034, India.\\
$^2$Centre for Theoretical Physics, Jamia Millia Islamia, New Delhi-110025,
India.}

\abst{ The two-photon ghost interference experiment,
generalized to the case of massive particles, is theoretically analyzed.
It is argued that the experiment is intimately connected to a double-slit
interference experiment where, which-path information exists. The reason
for not observing first order interference behind the double-slit, is clarified.
It is shown that the underlying mechanism for the appearance of ghost
interference is, the more familiar, quantum erasure.}

\begin{document}
\maketitle

\section{Introduction}

A puzzling experiment which gave a dramatic demonstration of the nonlocal 
nature of quantum
correlations that exist in spatially separated entangled particles,
was reported by Strekalov et al.\cite{ghost}, and has come to be known
as ghost interference. 
In brief, the experiment goes as follows. 
A Spontaneous Parametric Down-Conversion (SPDC) source S sends out pairs of
two entangled photons, which we call photon 1
and photon 2 (see Fig. 1). A double-slit is placed in the path 
of photon 1.

\begin{figure}
\centerline{\resizebox{10.0cm}{!}{\includegraphics{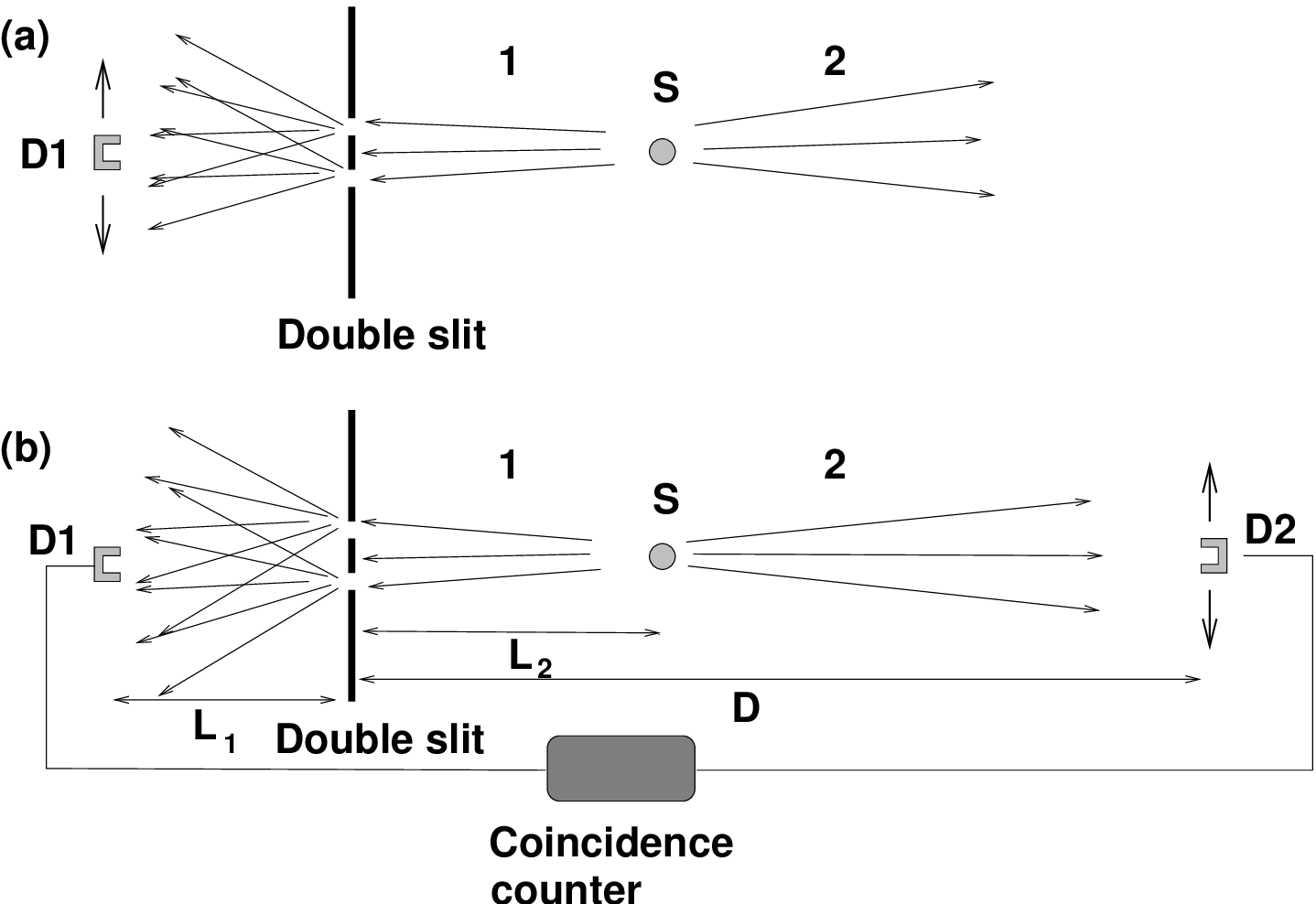}}}
\caption{ An SPDC source generates photon pairs - one goes left,
and the other right. (a) Putting a double slit in the path of photon 1 results
in no interference. (b)  Counting of photon 2 in coincidence with a
{\em fixed} detector D1 clicking, results in a ghost interference.}
\end{figure}
The results of the experiment are as follows.

\noindent ({\it a}) When photons 1 are detected 
using a detector placed behind the double-slit, no first order interference 
is observed for photon 1. This is surprising because one would have 
expected interference to be observed due to the double-slit in the path of 
photons 1. For
photons 2, first order interference is neither expected, nor is it seen.

\noindent ({\it b}) When photons 2 are detected {\em in coincidence
with a fixed detector behind the double slit registering photon 1}, an
interference
pattern which is very similar to a double-slit interference pattern is 
observed,  
even though there is no double-slit in the path of photon 2. Changing the
position of the fixed detector does not change the interference pattern, but
only shifts it. 

Another curious thing is that the interference pattern is the same as what
one would
observe if one were to replace the lone photon 1 detector behind the 
double slit, with a source of light, and the SPDC source were absent. In 
other words, 
the standard Young's double slit interference formula works, if the distance
is taken to be the distance between the screen (detector) on which photon 2
registers, right through the SPDC source crystal, to the double slit. Photon
2 never passes through the region between the source S and the double 
slit. 

This experiment is aptly called ``ghost interference". Remarkable is the
fact that even though photon 2 never passes through the region between
the source S and the double slit, we see interference pattern for photon
2 as though a beam of photon 2, with source located at the position
of detector 1 is being spilt by the double-slit. This experiment has
become a subject of experimental and theoretical research attention
\cite{ghostimaging,rubin,zhai,jie}, and has been understood to be a
consequence of entanglement.  Zeilinger's group also independently
performed a ghost interference experiment using an optical grating
\cite{zeilinger}.

For explaining ghost interference, Strekalov et al. presented a geometrical model
which satisfactorily reproduces the observed pattern.  However, we believe
that the mechanism behind the emergence of ghost interference can be
understood better by looking at it from a different perspective. In this
paper, we present this new way of looking at ghost interference.
Strekalov et al. attribute the absence of first order interference in photon 1
to the large momentum spread of photon 1 - ``the `blurring out' of the
first order interference is due to the considerably large angular propagation
uncertainty of a single SPDC photon" \cite{ghost,ghostimaging}.
We will show that there is a more fundamental reason why a first order
interference can never be observed in an experiment with entangled photons.
In fact, we will show that the non-observation of the first order interference
for photon 1 is intimately related to the appearance of ghost interference
for photon 2.  We will also show that an interference for photon 1 can be
observed, under certain conditions. An experiment which is somewhat similar
in spirit, was carried out using electrons \cite{neder}. This indicates
that the phenomenon has to do with quantum correlations, and not with the
specific nature of particles involved.

\section{Theoretical analysis}

At the heart of this effect is the phenomenon of entanglement, which
applies as much to massive particles, as to photons.
For clarity, we will analyze the ghost interference experiment using
entangled particles, rather than photons. The results can easily be applied
to the case of photons.
Let there be two particles of identical mass, generated at the source S,
in an entangled state. We assume the entangled state to be of the
following form:
\begin{equation}
\Psi(y_1,y_2) = C\!\int_{-\infty}^\infty dp
e^{-p^2/4\sigma^2}e^{-ipy_2/\hbar} e^{i py_1/\hbar}
e^{-{(y_1+y_2)^2\over 4\Omega^2}}, \label{state}
\end{equation}
where $C$ is a normalization constant. The
$e^{-(y_1+y_2)^2/4\Omega^2}$ term is required so that the state (\ref{state}) 
is normalized in $y_1$ and $y_2$. 
This is a momentum entangled state, which is fairly general, barring the
use of Gaussian functions.
Integration over $p$ can be performed to obtain:
\begin{equation}
\Psi(y_1,y_2) = \sqrt{ {\sigma\over \pi\hbar\Omega}}
 e^{-(y_1-y_2)^2\sigma^2/\hbar^2}
e^{-(y_1+y_2)^2/4\Omega^2} .
\label{newstate}
\end{equation}
The physical meaning of the constants $\sigma$ and $\Omega$ will become clear
if we calculate the uncertainty in position and momentum of the two particles.
The uncertainty in momenta of the two particles is given by
\begin{equation}
\Delta p_{1y} = \Delta p_{2y} = 
        \sqrt{\sigma^2 + {\hbar^2\over 4\Omega^2}}. \label{dp}
\end{equation}
The position uncertainty of the two particles is given by
\begin{equation}
\Delta y_1 = \Delta y_2 = \sqrt{\Omega^2+\hbar^2/4\sigma^2}.
\end{equation}
So, now we know the position and momentum spread of both the particles in
this state. With time, the particles travel along the positive and
negative x-axis. The motion in the x-direction is disjoint from the
evolution in the y-direction, and is unaffected by entanglement. So,
in order to see the effect of the double slit on particle 1, we will
assume that state evolves for a time $t_0$ before particle 1 reaches the
double-slit.

The state of the entangled system, after this time evolution, can be
calculated using the Hamiltonian governing the time evolution, given by
\begin{equation}
\hat{H} = -{\hbar^2\over 2m} {\partial^2\over \partial y_1^2} 
            -{\hbar^2\over 2m} {\partial^2\over \partial y_2^2} \label{H}
\end{equation}
After a time $t_0$, (\ref{newstate}) assumes the form
\begin{equation}
\Psi(y_1,y_2,t_0) = {1\over \sqrt{{\pi\over 2}(\Omega+{i\hbar t_0\over m\Omega})
(\hbar/\sigma + {4i\hbar t_0\over m\hbar/\sigma})}}
\exp\left[{-(y_1-y_2)^2\over \hbar^2/\sigma^2  + {4i\hbar t_0\over m}}
\right]
\exp\left[{-(y_1+y_2)^2\over \left(\Omega^2  +
{i\hbar t_0\over m}\right)} \right] 
\label{Psit}
\end{equation}
We wish to point out that the use of (\ref{H}) is not an absolute necessity
for obtaining  the time evolution of the state. For example,  if one
considers the particle to be an envelope of waves , the time evolution can be
obtained easily. In that case, $\left({d^2\omega(k)\over dk^2}\right)_{k_0}$,
where $k_0$ is the wave-vector value where $\omega(k)$ peaks, plays the role
of $\hbar/m$.
The time evolution for a photon state can be obtained similarly \cite{mandel}.

\subsection{Double-slit and which-way information}

Imagine the slit to be a position filter - it allows portions of the
wavefunction in front of the slit, to go through. Let us assume that what
emerges from a slit is a localized Gaussian packet, whose width is the 
width of the slit. So, if the two slits are A and B, the packets which
pass through will be, say, $|\phi_A(y_1)\rangle$ and $|\phi_B(y_1)\rangle$,
respectively.

The entangled state at time $t_0$, $|\Psi(y_1,y_2,t_0)\rangle$, can then be expanded in
terms of
components parallel to $|\phi_A(y_1)\rangle$ and $|\phi_B(y_1)\rangle$,
and orthogonal to those. We can write
\begin{equation}
|\Psi(y_1,y_2,t_0)\rangle = |\phi_A\rangle\langle\phi_A|\Psi\rangle
+ |\phi_B\rangle\langle\phi_B|\Psi\rangle +
|\chi\rangle\langle\chi|\Psi\rangle . \label{slit}
\end{equation}
where $|\chi(y_1)\rangle$ represents rest of the states in the Hilbert
space, orthogonal to $|\phi_A(y_1)\rangle$ and $|\phi_B(y_1)\rangle$.
So, the states of particle 2 that one has to calculate are
\begin{eqnarray}
\psi_A(y_2) &=& \langle\phi_A(y_1)|\Psi(y_1,y_2,t_0)\rangle \nonumber\\
\psi_B(y_2) &=& \langle\phi_B(y_1)|\Psi(y_1,y_2,t_0)\rangle \nonumber\\
\psi_\chi(y_2) &=& \langle\chi(y_1)|\Psi(y_1,y_2,t_0)\rangle \label{psi}
\end{eqnarray}

So, the state we get after particle 1 crosses the double-slit is:
\begin{equation}
|\Psi(y_1,y_2)\rangle = |\phi_A\rangle|\psi_A\rangle
+ |\phi_B\rangle|\psi_B\rangle +
|\chi\rangle|\Psi_\chi\rangle ,
\end{equation}
where $|\phi_A\rangle$ and $|\phi_B\rangle$ are states of particle 1,
and $|\psi_A\rangle$ and $|\psi_B\rangle$ are states of particle 2.
The first two terms represent the amplitudes of particle 1 passing through
the slits, and the last term represents the amplitude of it getting
reflected/blocked.
Because of the linearity of Schr\"odinger equation, these two pieces of
the wavefunction will evolve independently, without affecting each
other. Because we are interested only in situations where particle 1 passes
through the slit, we might as well throw away the term which represents
particle 1 not passing through the slits. If we do that, we have to normalize
the remaining part of the wavefunction, which looks like
\begin{equation}
|\Psi(y_1,y_2)\rangle = {1\over C} (|\phi_A\rangle|\psi_A\rangle
+ |\phi_B\rangle|\psi_B\rangle),
\end{equation}
where $C = \sqrt{\langle\psi_A|\psi_A\rangle + \langle\psi_B|\psi_B\rangle}$.

In the following, we assume that $|\phi_A\rangle$, $|\phi_B\rangle$, are
Gaussian functions in space:
\begin{equation}
\phi_A(y_1) = {1\over(\pi/2)^{1/4}\sqrt{\epsilon}} e^{-(y_1-y_0)^2/\epsilon^2}
,~~~~~
\phi_B(y_1) = {1\over(\pi/2)^{1/4}\sqrt{\epsilon}} e^{-(y_1+y_0)^2/\epsilon^2},
\end{equation}
where $\pm y_0$ is the y-position of slit A and B, respectively, and $\epsilon$
their widths. Thus, the distance between the two slits is $2 y_0$.

Using (\ref{psi}) and (\ref{Psit}), wavefunctions $|\psi_A\rangle$,
$|\psi_B\rangle$ can be calculated, which, after normalization, have the form
\begin{equation}
\psi_A(y_2) = C_2 e^{-{(y_2 - y_0')^2 \over \Gamma^2}},~~~
\psi_B(y_2) = C_2 e^{-{(y_2 + y_0')^2 \over \Gamma^2}} ,
\end{equation}
where $C_2 = {1\over (\pi/2)^{1/4}\sqrt{\Gamma}}$,
\begin{equation}
y_0' = {y_0 \over {4\Omega^2\sigma^2/\hbar^2+1\over
4\Omega^2\sigma^2/\hbar^2-1} + {4\epsilon^2 \over 4\Omega^2-\hbar^2/\sigma^2}},
\end{equation}
and
\begin{equation}
\Gamma^2 = \frac{{\hbar^2\over\sigma^2}(1+{\epsilon^2+2i\hbar t_0/m\over 4\Omega^2})
 + \epsilon^2+2i\hbar t_0/m}{1 + {\epsilon^2+2i\hbar t_0/m\over\Omega^2} +
{\hbar^2\over 4\Omega^2\sigma^2}} +{2i\hbar t_0\over m} .
\end{equation}

The state which emerges from the double slit, now assumes the form
\begin{equation}
\Psi_r(y_1,y_2) = 
C_1 e^{-(y_1-y_0)^2/\epsilon^2}
C_2 e^{-{(y_2 - y_0')^2 \over \Gamma^2}}
+ C_1 e^{-(y_1+y_0)^2/\epsilon^2}
C_2 e^{-{(y_2 + y_0')^2 \over \Gamma^2}} \label{virtual},
\end{equation}
where $C_1 = \left({2\epsilon^2\over\pi}\right)^{1/4}$
Equation (\ref{virtual}) represents two wave-packets of particle 1,
of width $\epsilon$, and localized at $\pm y_0$, entangled with two
wave-packets of particle 2, of width
${\sqrt{2}|\Gamma|^2\over\sqrt{\Gamma^2+\Gamma^{*2}}}$, localized at
$\pm y_0'$.

The state
(\ref{virtual}) represents particle 1 passing through a double-slit. But
if $|\psi_A\rangle$ and $|\psi_B\rangle$ are orthogonal, 
the amplitudes of particle 1 passing through the two slits are correlated with
two distinguishable states of particle 2. Hence, in principle, a measurement
on particle 2 can reveal which slit particle 1 passed through. According
to the Complementarity principle, no interference can be observed in such
a situation. So, no interference can be seen in particle 1 because particle
2 carries the ``which-way" information about particle 1. This is the fundamental
reason for photon 1 not showing interference in the ghost interference
experiment, and not its large momentum spread.

\subsection{Entanglement and virtual double-slit}

From (\ref{virtual}) one can see that the state of particle 2 also involves
two spatially separated, localized Gaussians, correlated with states of
particle 1. So, because of entanglement, particle 2 also behaves as if it
has passed through a double-slit of separation $2y_0'$. In other words,
because of entanglement, particle 1 passing through the double-slit,
creates a {\em virtual double-slit} for particle 2. This view is in agreement
with the observed optical imaging by means of entangled photons \cite{imaging}.
It appears natural that
particle 2, passing through this virtual double-slit, should show an
interference pattern. However, this can happen only when the wave-packets
overlap, after evolving in time.

Before reaching detector D2, particle 2 evolves for a time $t$.
The time evolution, governed by (\ref{H}), transforms the state (\ref{virtual})
to
\begin{eqnarray}
\Psi_r(y_1,y_2,t) &=& 
C_1(t) e^{-{(y_1-y_0)^2\over\epsilon^2+2i\hbar t/m}}
C_2(t) e^{-{(y_2 - y_0')^2 \over \Gamma^2+2i\hbar t/m}}
\nonumber\\
&&+ C_1(t) e^{-{(y_1+y_0)^2\over\epsilon^2+2i\hbar t/m}}
C_2(t) e^{-{(y_2 + y_0')^2 \over \Gamma^2+2i\hbar t/m}}, \nonumber\\
\end{eqnarray}
where 
\begin{equation}
C_1(t) = {1\over (\pi/2)^{1/4}\sqrt{\epsilon+2i\hbar t/m\epsilon}},~~~
C_2(t) = {1\over (\pi/2)^{1/4}\sqrt{\Gamma+2i\hbar t/m\Gamma}}.
\end{equation}
Before proceeding further, we need to simplify the expression for $\Gamma$.
We assume the spatial extent of the wave-function $\Psi(y_1,y_2)$ to be
large, namely, $\Omega \gg \epsilon$ and $\Omega \gg \hbar/\sigma$. In this
limit,
\begin{equation}
\Gamma^2 \approx \gamma^2 + 4i\hbar t_0/m ,
\end{equation}
where $\gamma^2 = \epsilon^2 + \hbar^2/\sigma^2$ and $y_0' \approx y_0$.
We are now in a position to calculate the probability of finding particle 1
at $y_1$ and particle 2 at $y_2$. This is given by
\begin{eqnarray}
|\Psi_r(y_1,y_2,t)|^2 &=& 
|C_1(t)C_2(t)|^2\times \left( 
\exp\left[{-{2(y_1-y_0)^2\over\epsilon^2+({2\hbar t\over m\epsilon})^2}}
-{2(y_2 - y_0')^2 \over \gamma^2+({2\hbar (t+2t_0)\over m\gamma})^2}\right]\right. \nonumber\\
&&+ \exp\left[{-{2(y_1+y_0)^2\over\epsilon^2+({2\hbar t\over m\epsilon})^2}}
{-{2(y_2 + y_0')^2 \over \gamma^2+({2\hbar (t+2t_0)\over m\gamma})^2}}\right]
\nonumber\\
&&+ \left.\exp\left[{-{2(y_1^2+y_0^2)\over\epsilon^2+({2\hbar t\over m\epsilon})^2}
-{2(y_2^2 + y_0'^2) \over \gamma^2+({2\hbar (t+2t_0)\over m\gamma})^2}}\right] 
\times 2\cos\left[\theta_1 y_1 + \theta_2 y_2\right]\right), \label{pattern}
\nonumber\\
\end{eqnarray}
where
\begin{equation}
\theta_1 = {8y_0\hbar t/m\over \epsilon^4 + 4\hbar^2t^2/m^2},~~~
\theta_2 = {8y_0\hbar (t+2t_0)/m\over \gamma^4 + 4\hbar^2(t+2t_0)^2/m^2}.
\end{equation}
We can now make contact with the ghost interference experiment, where
detector D1 is kept fixed and detector D2 is scanned along the y-axis.
If we fix $y_1$, the cosine term in (\ref{pattern}) represents oscillations
as a function of $y_2$. This implies that if particle 2 is detected in
coincidence with particle 1 being detected at a fixed position $y_1$, then
it shows interference. This is ghost interference. In the expression for
$\theta_2$, $\gamma$ is a measure of the width
of the virtual slits created because of particle 1 passing through the
double-slit. So, it is clear that the ghost interference is an effect
due to the virtual slits formed for particle 2, because of a measurement on
the spatially separated particle 1, with which it is correlated due to 
entanglement.

\subsection{Erasing the which-way information}

Coming back to ghost interference, one might wonder that if the virtual
slits are created by particle 1 passing through the double-slit, what is
the need for fixing detector D1 and doing a coincident count. The
answer is, particle 1 carries which-way information about particle 2.
Meaning, particle 1 can be potentially detected in such a manner that would
tell us
which virtual slit, A or B, did particle 2 pass through. And Bohr's
complementarity principle tells us that in such a situation, interference
cannot be observed \cite{bohr}. By detecting particle 1 at a fixed position, where
contributions from both the slits are present, we add the two contributions,
and thus {\em erase} the information about which slit particle 1, and 
particle 2,
passed through. Once the which-way information is erased, the interference
can come back, and it does. So, the mechanism behind the appearance of
ghost interference is two-fold. Quantum entanglement of the two particles leads
to the creation of a virtual double-slit for particle 2, and
{\em quantum erasure} \cite{eraser} of the which-way information (again
via entanglement) leads to
appearance of interference from the virtual double-slit.

\begin{figure}[h]
\centerline{\resizebox{10.0cm}{!}{\includegraphics{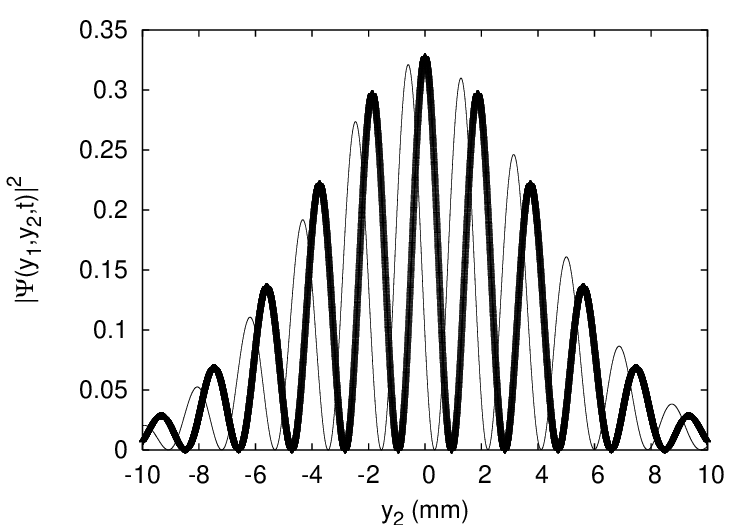}}}
\caption{ Probability density of particle 2 as a function of the position of
detector D2, for $\lambda_d=314$ nm, $D=3$ m, $L_1=1$ m, $2y_0=0.5$ mm and
$\epsilon=0.05$ mm. The dark pattern corresponds to $y_1=0$ mm, and the lighter
pattern corresponds to $y_1=0.2$ mm.}
\end{figure}

Scully Englert and Walther proposed a setup for quantum eraser, where
the which-way detector is a two-state system \cite{scully}. Quantum
erasure is performed when the particles are detected in coincidence
with one of the two special states of the which-way detector, which
do not discriminate between the two paths of the particle. Corresponding
to these two states of the which-way detector, two interference patterns
are obtained which are complementary, meaning, they add up to give no
interference. In the case of ghost interference, the role of the 
states carrying which-way information is played by the position of particle 1,
namely $y_1$, which is a continuous variable. Detection of particle 1
by a {\em fixed} detector D1 destroys the which-way information carried
by it. Thus D1 acts as an eraser of the which-way information.
From (\ref{pattern}), one can see that for a fixed $y_1$, the term
$\theta_1 y_1$ acts as an extra phase for the cosine function in $y_2$.
Thus, the whole interference pattern is shifted, depending on the position
of D1 (see Fig. 2), 
so that when all the D1 positions are added, it results in the destruction of
the interference pattern. This is the reason why, in Strekalov et al.'s
experiment, no interference for photon 2 is observed without coincident
counting with a fixed D1.

\subsection{Where is the virtual slit located?}

One must have already noticed something strange about (\ref{pattern}),
namely, in the terms for particle 2, the time which appears is $t+2t_0$
as opposed to just $t$ for particle 1. In the actual experiment, time is
not what is the meaningful quantity - it is the distance the particle
travels, that is relevant. Let us translate our results to the situation
where one just measures the distance.  For that we assume that particle 2
travels along the x-axis, with a momentum $p$. In time $t_0$, both particle
1 and particle 2 travel a distance $L_2$. During time $t$, particle 1
travels a distance $L_1$ to reach D1, and particle 2 travels the same
distance to reach D2. So, the time $t+2t_0$ corresponds to the distance
between the double-slit and D2, that is, $D$. Using this strategy, we can
write $\hbar (t+2t_0)/m = \hbar v(t+2t_0)/p = \lambda_d v(t+2t_0)/2\pi =
\lambda_d D/2\pi$, where
$\lambda_d$ is the d'Broglie wavelength of the particle and $v$ its velocity.
The expression $\lambda_d D/2\pi$ will also hold for a
photon provided, one uses the wavelength of the photon in place of
$\lambda_d$. The probability of coincident click of D1 and D2 is given by
$P(y_1,y_2) = |\Psi_r(y_1,y_2,t)|^2$, which has the following form
\begin{eqnarray}
P(y_1,y_2) &=& |C_1(t)C_2(t)|^2 \left(  
\exp\left[-{2(y_1-y_0)^2\over\epsilon^2+(\lambda_d L_1/\pi\epsilon)^2}
-{2(y_2 - y_0)^2 \over \gamma^2+(\lambda_d D/\pi\gamma)^2}\right]\right. \nonumber\\
&+& \exp\left[-{2(y_1+y_0)^2\over\epsilon^2+(\lambda_d L_1/m\epsilon)^2}
-{2(y_2 + y_0)^2 \over \gamma^2+(\lambda_d D/\pi\gamma)^2}\right]
\nonumber\\
&+& \exp\left[-{2(y_1^2+y_0^2)\over\epsilon^2+(\lambda_d L_1/\pi\epsilon)^2}
-{2(y_2^2 + y_0^2) \over \gamma^2+(\lambda_d D/\pi\gamma)^2}\right] 
\times 2\cos\left[\theta_1 y_1 + \theta_2 y_2\right]\left.\right), \label{plot}
\nonumber\\
\end{eqnarray}
where $\theta_1 = {4y_0\lambda_d L_1/\pi\over \epsilon^4 + \lambda_d^2L_1^2/\pi^2}$,
$\theta_2 = {4y_0\lambda_d D/\pi\over \gamma^4 + \lambda_d^2D^2/\pi^2}$.
Equation (\ref{plot}) tells us that the fringe width of the pattern for particle 2 is given by
\begin{equation}
w_2 = {2\pi \over \theta_2} = 2\pi {\lambda_d^2 D^2/4\pi^2 + \gamma^4/4\over
       2y_0 \lambda_d D/2\pi} = {\lambda_d D\over 2y_0}+
{\gamma^4\pi\over 4y_0\lambda_d D}
\end{equation}
For $\gamma^2 \ll \lambda_d D$, we get the familiar Young's double-slit
interference formula,
\begin{equation}
w_2 \approx {\lambda_d D\over 2y_0},
\end{equation}
where $2y_0$ is the separation between the slits. Notice that $D$ is the
strange distance from the detector D2, right through the source, to the double
slit (see Fig. 1). Particle 2 never passes through the region between the
source and the
double-slit. This is exactly what was observed in Strekalov et al.'s experiment.
Although the virtual double-slit for particle 2 comes into being only after
particle 2 travels a distance $L_2$ from the source, the particle carries
with itself the phase information of its evolution from the source for the
time $t_0$. Because of coincident counting, the change in phase because of
the evolution of particle 1 is added to that of particle 2, and {\em it
appears as if} particle 2 traveled a distance $2L_2$, which is double the
actually traveled distance.  So, we
see that although the virtual double-slit comes into being after particle
1 enters the real double-slit, for all practical purposes, {\em it appears
as if} the virtual double-slit is located exactly at the real double-slit,
{\em behind the source}. We should also mention that
for values of various parameters corresponding to Strekalov et al.'s experiment,
their results are faithfully reproduced by (\ref{plot}).

\subsection{Interference for particle 1}

Eqn. (\ref{plot}) is reasonably symmetric in $y_1$ and $y_2$, except for
the difference in the widths of the real and virtual slits, and $L_1$
appearing for particle 1, and $D$ appearing for particle 2. It is but
natural to expect that by fixing D2 at some $y_2$, and counting particle 1
in coincidence with D2, should show
an interference. The fringe width of the interference pattern is given by
\begin{equation}
w_1 = {2\pi \over \theta_1} = 2\pi {\lambda_d^2 L_1^2/4\pi^2 + \epsilon^4/4\over
       2y_0 \lambda_d L_1/2\pi} \approx {\lambda_d L_1\over 2y_0}
\end{equation}
The fringe width is exactly what one would expect from a conventional first
order interference. In this sense, this pattern is not as spectacular as that
for particle 2. The term $\theta_2 y_2$ now acts as an additional phase of
the cosine, and this leads to a shift in the interference pattern for
particle 1 if the position of D2 is changed.

Kim et al. \cite{kim} performed an experiment which, in the context of the
experiment described here, would amount to looking for interference in
particle 1, by erasing the which-way information carried by particle 2.
They demonstrate that the two-slit interference is recovered once the which-way
information is erased. However, they go further than that.
They not only actually acquire which-way information for each passing
particle, which does not happen in the experiment described here, they
demonstrate that which-way information can be erased much after the
particle has been physically detected, and still leads to recovery
of interference. The physical interpretation of such ``delayed-choice" quantum
erasure has been much debated upon \cite{wheeler,esw,mohrhoff}.

\section{Conclusion}

From the preceding analysis, we conclude that in the ghost interference
experiment, the reason for the absence of first order interference for
particle 1 is that the which-way information for particle 1, is carried
by particle 2. By complementarity, no interference can be observed in
such a situation, in principle.
Particle 2 can show interference because, by virtue of entanglement,
it experiences a
virtual double-slit due to particle 1 passing through the double-slit.
However, particle 1 carries information on which virtual slit particle 2
passed through, and that
washes out any potential interference. By fixing D1 and doing a coincident
count of particle 2, one is erasing the which-way information.
This quantum erasure leads to the appearance of ghost interference in particle
2. A corollary of the result is that particle 1 can also show interference if
it is detected in coincidence with a fixed D2.
The general analysis presented here shows that ghost interference can be
observed for entangled massive particles too.

\section*{Acknowledgments}
Authors thank Pankaj Sharan for useful discussions.


\begin{thebibliography}{0}

\bibitem{ghost} D. V. Strekalov, A. V. Sergienko, D. N. Klyshko, and Y. H. Shih,
{\em Phys. Rev. Lett.} {\bf 74} (1995), 3600.

\bibitem{ghostimaging} M. D'Angelo, Y-H. Kim, S. P. Kulik and Y. Shih,
{\em Phys. Rev. Lett.} {\bf 92} (2004), 233601.

\bibitem{rubin} S. Thanvanthri and M. H. Rubin,
{\em Phys. Rev. A} {\bf 70} (2004), 063811.

\bibitem{zhai} Y-H. Zhai, X-H. Chen, D. Zhang, L-A. Wu,
\textit{Phys. Rev. A} {\bf 72} (2005), 043805.

\bibitem{jie} L. Jie, C. Jing,
\textit{Chinese Phys. Lett.} {\bf 28} (2011), 094203

\bibitem{zeilinger} R. Christanell, W. Weinfurter, and A. Zeilinger,
{\em The Technical Digest of the European Quantum Electronic Conference,
EQEC`93}, Florence, 1993 (unpublished), p. 872.

\bibitem{neder} I. Neder, M. Heiblum, D. Mahalu, V. Umansky,
{\em Phys. Rev. Lett.} {\bf 98} (2007), 036803.

\bibitem{mandel} L. Mandel and E. Wolf, \textit{Optical Coherence and Quantum
Optics} (Cambridge, University Press, 1995).

\bibitem{imaging} T.B. Pittman, Y.H. Shih, D.V. Strekalov, A.V. Sergienko,
{\em Phys. Rev. A}, {\bf 52} (1995), R3429.

\bibitem{bohr} N. Bohr, {\em Nature} {\bf 121} (1928), 580.

\bibitem{eraser} M. O. Scully and K. Dr\"{u}hl, 
{\em Phys. Rev. A} {\bf 25} (1982), 2208.

\bibitem{scully} M. O. Scully, B.-G. Englert and H. Walther,
{\em Nature (London)} {\bf 351} (1991), 111.

\bibitem{kim} Y-H Kim, R. Yu, S.P. Kulik, Y. Shih, M.O. Scully,
{\em Phys. Rev. Lett.} {\bf 84} (2000), 1-5.

\bibitem{wheeler} J.A. Wheeler, pp. 182-213 in {\em Quantum Theory and
Measurement}, J.A. Wheeler and W.H. Zurek edit.,
(Princeton University Press, 1984).

\bibitem{esw} B.-G. Englert, M. O. Scully and H. Walther, {\em Am. J. Phys.}
{\bf 67}, 325 (1999).

\bibitem{mohrhoff} U. Mohrhoff, {\em Am. J. Phys.} {\bf 67}, 330 (1999).

\end{thebibliography}
\end{document}